\documentclass[useAMS,usegraphicx,usenatbib]{mn2e}

 \long\def\symbolfootnote[#1]#2{\begingroup%
 \def\thefootnote{\fnsymbol{footnote}}\footnote[#1]{#2}\endgroup}

\title[Are z$\sim$5 QSOs found in the most massive high redshift over-densities?]{Are z$\sim$5 QSOs found in the most massive high redshift over-densities? }
\author[K. Husband et al.]{K. Husband $^1$, M.~N. Bremer $^1$, E.~R. Stanway $^2$, L.~J.~M. Davies $^1$, M.~D. Lehnert $^{3,4}$,\and L.~S. Douglas $^3$ \\
$^1$H.H.~Wills Physics Laboratory, University of Bristol, Tyndall Avenue, Bristol, BS8 1TL, UK. \\
$^2$Department of Physics, University of Warwick, Gibbet Hill Road, Coventry, CV4 7AL, UK. \\
$^3$GEPI, Observatoire de Paris, CNRS, Universit\'e Paris Diderot, 5 place Jules Janssen, 92190 Meudon, France. \\
$^4$Visting Astronomer, Institut d'Astrophysique de Paris, 98 bis boulevard Arago, 75014 Paris, France.} 

\begin{document}
\maketitle

\begin{abstract}
Luminous high-redshift QSOs are thought to exist within the most massive dark matter haloes  in the young Universe. As a consequence they are likely to be markers for biased, over-dense regions where early galaxies cluster, regions that eventually grow into the groups and clusters seen in the lower-redshift universe. In this paper we explore the clustering of galaxies around $z\sim 5$ QSOs as traced by Lyman break Galaxies (LBGs). We target the fields of three QSOs using the same optical imaging and spectroscopy techniques used in the ESO Remote Galaxy Survey, (ERGS, \citealp{Douglas09,Douglas10}), which was successful in identifying individual clustered structures of LBGs. We use the statistics of the redshift clustering in ERGS to show that two of the three fields show significant clustering of LBGs at the QSO redshifts. Neither of these fields is obviously over-dense in LBGs from the imaging alone; a possible reason why previous imaging-only studies of high redshift QSO environments have given ambiguous results. This result shows that luminous QSOs at $z\sim 5$ are typically found in over-dense regions. The richest QSO field contains at least nine spectroscopically confirmed objects at the same redshift including the QSO itself, seven LBGs and a second fainter QSO. While this is a very strong observational signal of clustering at $z\sim 5$, it is of similar strength to that seen in two structures identified in the `blank sky' ERGS fields. This indicates that, while over-dense, the QSO environments are not more extreme than other structures that can be identified at these redshifts. The three richest structures discovered in this work and in ERGS have properties consistent with that expected for proto-clusters and likely represent the early stages in the build-up of massive current-day groups and clusters.
\end{abstract}


%%%%%%%%%%%%%%%%%%%%%%%%%%%%%%%%% Intro %%%%%%%%%%%%%%%%%%%%%%%%%%%%%%%%%%%%%%%%%%%%%%%%%%%%%%%%%%%%%%%%%%%%
\section{Introduction}
\label{secintro}
\symbolfootnote[0]{Based on observations made at the European Southern Observatory Very Large Telescope, Paranal, 
Chile (ESO programme numbers 082.A-0354(A), 083.A-0618(A), 084.A-0251(B), 085.A-0444(B) \& 087.A-0589(A)(B)).}

Our theoretical understanding of the evolution of structure appears to be well advanced and has been thoroughly explored through large numerical simulations \citep[e.g.][]{Springel05}. Such simulations, using the current cosmological parameters determined from observations of distant supernovae, the microwave background and other tracers, can largely predict the structures that we see in observational studies of the relatively nearby universe.

By contrast, there are currently few observations that directly detail the early growth of structures such as groups and clusters in the first few Gyrs. Massive clusters can be directly observed out to $z\sim 2$  from wide-area optical/IR and X-ray surveys \citep[e.g.][]{Bremer06,Fassbender11,Santos11,Spitler12,Stanford12,Zeimann12} and proto-clusters to somewhat higher redshifts by targeting either powerful radio sources \citep[e.g.][]{Miley04,Venemans07,Hatch11, Kuiper11, Kuiper12} or submm sources \citep[e.g.][]{Capak11}. At the highest redshifts we have detected very few systems that we can be confident are the progenitors of the massive structures that exist in the current-day Universe \citep[e.g.][]{Ota08,Toshikawa12}. There is a need to identify and study these individual systems back to the earliest stages in their evolution, in order to understand the interplay between the formation and evolution of these structures and the galaxies that they contain.

While it is possible to identify clustered regions at the highest redshifts through spectroscopy of randomly-selected deep fields, few have been identified in this manner at $z>4$ [the only known z $>$ 4 structures are the two in \citet[][hereafter D10]{Douglas10} at $z\sim 5$, the $z\sim 5.7$ structures in \citet{Ouchi05} and the $z\sim 6$ structure in \citet{Toshikawa12}]. Carrying out these `blind' studies is potentially observationally expensive, given the volumes and therefore number of objects which need to be targeted. Clearly, it is beneficial to have more effective methods for identifying potential high redshift clustering prior to large-scale observational programs.

It has long been thought that luminous high-redshift QSOs should be found in the most massive hosts and haloes \citep{Turner91} and that these systems should be associated with over-densities in the early matter distribution \citep{Efstathiou88}. Subsequent analytical and computational modelling indicates powerful high-redshift QSOs, hosted in massive dark matter haloes, may trace some of the most over-dense regions in the early Universe, and that these regions are likely to be the progenitors of massive low-redshift clusters \citep{Springel05,Li07,Trenti07,Matteo08,Angulo12}. The most recent of these works predict that the current-day descendants of structures hosting luminous high-redshift QSOs have a range of masses from $\sim 6\times 10^{13}$M$_\odot$  to more than $10^{15}$M$_\odot$ (i.e. the most massive low-z clusters), with the typical mass approaching $10^{15}$M$_\odot$. 

Prompted in part by these predictions, several observational studies \citep{Stiavelli05,Zheng06,Kim09,Utsumi10} have probed the environments of $z\ge 5$ QSOs for galaxies at the same redshift, in order to identify present day clusters in the earliest stages of formation. These studies have resulted in an ambiguous picture, with four of the QSO fields studied showing evidence for clustering and three showing no evidence for significant over-densities. In part this is due to the methods used. Many of the studies have targeted $z\ge 6$ QSOs  and then searched for over-densities in the number counts of dropout or Lyman break galaxies (LBGs), using photometric data with little or no spectroscopy.  While the LBG technique can, and does, select galaxies out to the highest redshifts, due to the increasing luminosity distance, brighter ground-based sky against which to select and the rapidly-evolving luminosity function with redshift \citep{Bouwens11,Lorenzoni12}, it is significantly more challenging to work at $z\sim 6$ than at $z\sim 5$ - despite the difference in look-back time being just $\sim 240$ Myr. Given the variation in number counts at faint levels and on arcminute scales, searching for an excess of sources in the near-environment of a QSO is likely to be a blunt test of clustering of galaxies at the QSO redshift. The broad-band drop-out selection typically identifies sources over a range of redshifts ($\Delta z \sim 0.5-1$) and any clustering is likely to be on scales of $\Delta z \sim 0.1$ or less. Thus, an over-density would have to be sufficiently strong to be reliably detected as a perturbation in the galaxy number counts.  Given that the surface density of $z\sim 5-6$ drop-outs seen in deep ground-based data is typically 1 per few arcmin$^2$ \citep[e.g.][]{Lehnert03}, an excess of just a handful of objects within $\Delta z \sim 0.1$ of a QSO's redshift is a significant signature of clustering. However, this signal would be lost in the background number counts without comprehensive spectroscopic follow up.

In this work we target the fields of three $z\sim 5$ QSOs, using an identical approach to that which we used in the ESO Remote Galaxy Survey \citep[ERGS;][]{Douglas09, Douglas10}.  In ERGS we spectroscopically confirmed 70 z $\sim$ 5 LBGs in 10 $\sim45$ arcmin$^2$, non-QSO fields using VLT/FORS2 \citep[][D10]{Douglas09}. While the majority of the ERGS fields showed no evidence for redshift clustering (typically there were zero or one LBG in each $\Delta z=0.1$ bin between $4.6<z<5.6$), two fields contained `spikes' in their redshift distribution indicating significant clustering of galaxies into structures. The multi-wavelength properties of those galaxies and structures have been probed in several papers \citep{Stanway08,Stanway10,Davies10,Davies12}. The purpose of this paper is to make a fair and robust comparison between fields which are either centered on a luminous QSO or not. By using a similar observing strategy for the QSO fields as in ERGS, not only can we be sure that if these QSOs are found at the centre of over-densities of LBGs then we will identify this large scale structure, but we can also compare the strength and character of any clustering to that seen in the `blank' sky fields of ERGS.

All optical magnitudes in this paper are quoted in the AB system \citep{Oke83} and a $\Lambda$CDM cosmology with H$_{0}$=71 kms$^{-1}$Mpc$^{-1}$, $\Omega_{M}$=0.3 and $\Omega_{\Lambda}$=0.7 is used throughout.

%%%%%%%%%%%%%%%%%%%%%%%%%%%%%%%%%%%%%%%%%%%%%%%%%%%% Imaging %%%%%%%%%%%%%%%%%%%%%%%%%%%%%%%%%%%%%%%%%%%%%%%%%%%%%%%%%%%%%%%%%%%%
\section{Object Selection and Observations}

The three QSOs targeted in this study were chosen from the $\sim 20$ SDSS QSOs with Decl.~$<10^\circ$ and spectroscopically-confirmed redshifts of between $4.9<z<5.2$ prior to 2009.  The three objects were selected to give an available target at any time during the year, i.e.~to be $\sim 8$ hours apart in RA, and to have a small range in UV luminosity of no more than an order of magnitude. The redshift range was chosen to match the most sensitive region of the ERGS selection function (see fig.~7 of D10). All three of the quasars chosen are very luminous with absolute magnitudes of between -28.1 to -29.5. One of these fields (J0338+0021) has a second, less-luminous, non-SDSS QSO at a similar redshift in the same FORS2 field of view \citep{Djorgovski03}. However, this was not taken into account when selecting this field.

Three $\sim$ 7\arcmin\ $\times$ 7\arcmin\ regions centered on the QSOs were imaged in $R,I$ and $z$ using FORS2/VLT and subsequently reduced in an analogous manner to the ERGS fields (see Table \ref{summary_images}).  Sextractor \citep{Bertin96} was used in dual mode to create $I$-band selected catalogues of objects in each of the fields. $z\sim 5$ LBG candidates were then selected using the same techniques and colour selection criteria as those used in D10, but with minor alterations intended to increase the efficiency of targeting galaxies at the QSO redshift.  A primary cut of $R-I>1.3$ was applied to select LBG candidates as, in the earlier work, this proved to be successful at identifying $z\sim5$ galaxies (fig.~7 \& 8 of D10). Given the uncertainties in the photometry, some objects at $4.9<z<5.0$ may be missed using this cut, especially if the object has significant Ly$\alpha$ emission contributing to the $R-$band flux \citep[see][for a discussion]{Stanway08}. One of the QSOs, J2130+0026 at $z=4.950$, has a colour in the FORS2 filters of $R-I=0.9$ (see Table 1) due to Ly$\alpha$+NV emission contributing to its $R$-band emission (an observed equivalent width of 2020 \AA). While LBGs do not have such large equivalent widths (see fig.~12 in D10), there is some chance that objects in this field, at the redshift of the QSO, could be missed by the primary $R-I$ selection. Therefore, our colour constraint was relaxed to include bluer objects with $R-I>1.0$.

The multi-colour images of each candidate object were then examined by eye and, because of our limited number of spectroscopic masks,  prioritised  for potential inclusion following the prescription outlined in D10. As in the previous work, we avoided using an explicit $I-z$ cut. However, because we found a trend of increasing $I-z$ colour with redshift between $4.6<z<5.6$ in ERGS (as expected), and in this work are specifically searching for galaxies at the lower end of this range, we gave higher priority to objects with $I-z<0.5$. Approximately 20 per cent of the spectroscopically-confirmed ERGS galaxies at $4.8<z<5.2$ were redder than this. Hence, such systems were not rejected outright.  The priority assigned to an object was also dependent whether or not its (ground-based) morphology was consistent with those of previously confirmed LBGs (unresolved or barely resolved) and  how close its colours were to the stellar locus (objects brighter than $I\sim 25$ and red $R-I$ and $I-z$ colours are almost invariably stars and low redshift galaxies). We also considered whether or not an object's photometry was uncompromised by crowding, confusion or scattered light from a nearby brighter object. The highest priority objects were classified as either priority 1 or 2, the first category indicating that the photometry was consistent with the source being at $z\sim 5$ and that there appeared to be no issues with the photometric measurements. Those objects that had $I-z >$ 0.5 or were brighter than typical  $z\sim 5$ galaxies were assigned as priority 2.

Two spectroscopic masks were created for each field. As with D10, sources were placed on the masks with the higher priority object having preference where dispersed spectra of two or more objects contended for the same detector area.  In general, there appeared to be little obvious 2-dimensional spatial clustering of the high priority targets, which would compromise the number of these sources placed on the masks through slit contention. Each mask was observed for 12800 seconds split into 20 exposures dithered along the slit, using the 300I grism and the OG590 order sorting filter. Approximately sixty objects were targeted in each field - the numbers of priority 1 and 2 objects targeted in each case are given in Table \ref{number}.

\begin{table*}
    \caption{The exposure times, seeing and 2$\sigma$ magnitude limits for the three QSO fields.}
    \centering

    \begin{tabular}[t]{c c c c c}
     Field & Filter & Total Exposure /hrs & Average Seeing /\arcsec & 2$\sigma$ Magnitude Limits \\
    \hline
J0338+0021   & R Special & 1.61 (20x290s) & 0.76 & 27.5 \\
\texttt{"}   & I Bessel  & 1.60 (24x240s) & 0.71 & 26.9 \\
\texttt{"}   & z Gunn    & 2.17 (65x120s) & 0.65 & 25.9 \\
J1204-0021   & R Special & 2.09 (26x290s) & 0.85 & 27.6 \\
\texttt{"}   & I Bessel  & 1.60 (24x240s) & 0.75 & 27.1 \\
\texttt{"}   & z Gunn    & 2.00 (60x120s) & 0.67 & 26.2 \\
J2130+0026   & R Special & 1.53 (19x290s) & 0.78 & 27.8 \\
\texttt{"}   & I Bessel  & 1.60 (24x240s) & 0.85 & 27.1 \\
\texttt{"}   & z Gunn    & 2.33 (70x120s) & 0.65 & 25.8 \\
    \end{tabular}
    \label{summary_images}
    \end{table*}

\begin{table*}
   \caption{The number of priority 1 and priority 2 objects in each field and the fraction of these observed. Overall a high proportion of the total number of priority 1 and priority 2 targets were observed. } 
    \centering
    \begin{tabular}[t]{c c c c c}
    Field & Priority 1 & Priority 1 & Priority 2 & Priority 2 \\
          & targets    & on mask    & targets    & on mask \\
    \hline
    J0338 & 42 & 20 (48\%) & 32 & 13 (41\%) \\
    J1204 & 24 & 18 (75\%) & 26 & 15 (58\%)\\
    J2130 & 27 & 12 (44\%) & 46 & 15 (33\%)\\
    \hline
    All   & 93 & 50 (54\%) &104 & 43 (41\%) \\
    \end{tabular}

    \label{number}
    \end{table*}

%%%%%%%%%%%%%%%%%%%%%%%%%%%%%%%%%%%%%%%%%%%%% Spectroscopy %%%%%%%%%%%%%%%%%%%%%%%%%%%%%%%%%%%%%%%%%%%%%%%%%%%%%%%%%%%%%%%%%%%%%
\section{Results}

Each spectrum was visually inspected by four of the authors (KH, MNB, LJMD and ERS) in order to classify the target and/or determine its redshift. For those objects identified as LBGs, we used the same classification as in D10, defined as follows. Those with a clear continuum break and/or emission line indicating a precise redshift were denoted grade  A. The resolution of each spectrum is $\sim 400$ kms$^{-1}$ at z=0, but in regions where a continuum break  fell in spectral regions containing strong sky-line residuals, the break could only be located to an accuracy of $\sim 1000$ kms$^{-1}$ at z=0. LBGs with no emission lines but with breaks falling in such regions were denoted grade B.

\begin{table*}
    \caption{The number of confirmed z$\sim$5 LBGs, low redshift objects and unknown sources due to low signal-to-noise spectra. Low-redshift objects have either clear continuum across the whole spectrum, an undulating spectrum indicative of a M-star or two or more emission lines indicating a low-redshift galaxy. Objects whose spectrum did not show evidence of being at low or high redshift were classed as unknown. Generally the higher priority objects were found to be LBGs and the lower priority objects to be low-redshift objects but there are also many priority 1 objects not confirmed as z$\sim$5 LBG, due to the low signal-to-noise spectra of faint sources making it difficult to see the continuum or be conclusive about a possible break.  The higher priority objects are generally fainter, as they were selected to be at high redshift and hence their spectroscopy is more likely to have low signal-to-noise. }
 
    \centering
    \begin{tabular}[t]{c c c c c}
    Sample & Confirmed high z & Low z & Unknown (low S/N) & Total \\
    \hline
    Pri 1 & 6 (+2 QSOs) (18\%) & 16 (36\%) & 21 (46\%) & 43 (+2 QSOs) \\
    Pri 2 & 4 (+1 QSO) (7\%) & 24 (42\%) & 29 (51\%) & 57 (+1 QSO) \\

    Pri 3 & 1  (4\%) & 19 (68\%) & 8  (28\%) & 28 \\ 
    Pri 4 & 0  (0\%) & 34 (79\%) & 9  (21\%) & 43 \\
    \end{tabular}
    \label{confirmations}
    \end{table*}

\begin{figure}
\centering
\includegraphics[width=0.95\columnwidth]{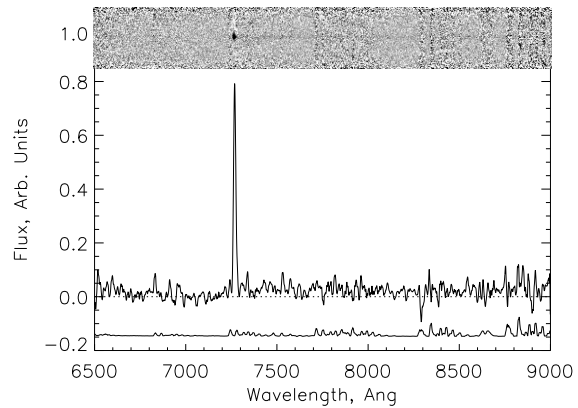}
\caption{An example LBG spectra with strong Lyman alpha emission giving a definite redshift (grade A). 
Note that the noisy vertical lines in the 2D plot are the residuals from subtracting the OH- emission 
from the sky. The sky lines are plotted underneath for comparison.}
\label{examplespectra}
\end{figure}

As well as those objects spectroscopically confirmed as LBGs, the
spectroscopy identified a similar range of  contaminating objects as
in our previous work, including low and intermediate ($z<1.5$)
redshift galaxies and M stars. Those objects resulting in spectra with
too low S/N to be identified as either LBGs or lower redshift objects
were classified as `unknown' (see Table \ref{confirmations}). An
example of a LBG with a Grade A  redshift from clear Lyman alpha
emission is shown in Fig.~\ref{examplespectra}. Results of our
spectroscopy generally confirm our priority ranking and show
contaminant discrimination is increasingly difficult from photometry
alone for fainter objects. Of the highest ranking objects, $\sim$20
per cent are spectroscopically confirmed as LBGs (see Table
\ref{confirmations}). A significant fraction of the other priority 1
and 2 sources have no clear spectroscopic classification.  In the
following we discuss our results field-by-field. 

\subsection{J0338+0021}
\begin{figure}
\includegraphics[width=.95\columnwidth]{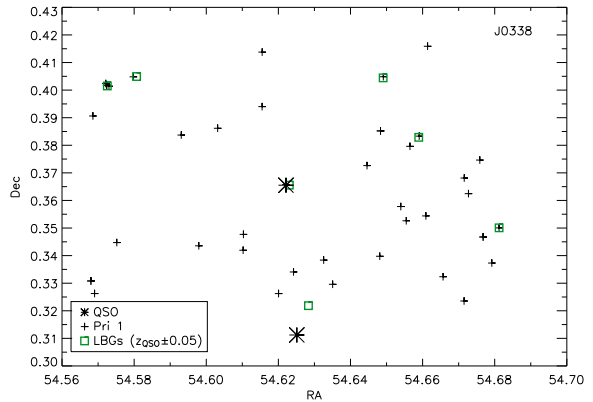}
\caption{The 2D spatial distribution of LBGs relative to the J0338+0021 QSO (star). There is also a second, less 
luminous QSO near the edge of the field (star). The LBGs within $\Delta z$ = 0.05 of the SDSS QSO are shown with squares 
and the other priority 1 targets from photometry are shown in the background as crosses. }
\label{j0338_pos}
\end{figure}

\begin{figure}
\includegraphics[width=.95\columnwidth]{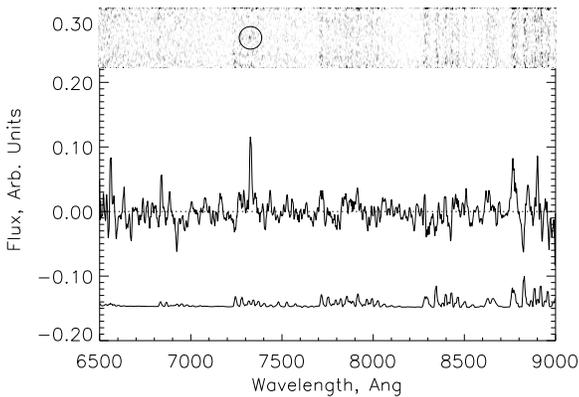}
\caption{The 2D and 1D spectra of the close companion to the J0338 QSO. The 2D spectrum is smoothed 
by the 2 by 2 pixel boxcar average and the possible Lyman alpha emission is circled. Again the sky lines 
are plotted underneath for comparison. At an $I$ magnitude of $\sim$26, this source is too faint for any 
continuum to be detected.}
\label{2a1spectra}
\end{figure}

Our observations confirm six LBGs (four grade A, two grade B) in the field of J0338+0021. Five of these cluster within $\Delta$z $<$ 0.05 ($\la$ 2500 kms$^{-1}$) of the QSO redshift, with a sixth at $\Delta z =$ 0.06. From the ERGS survey, we would expect typically zero or one object in a $\Delta z =$ 0.1 bin at this redshift in an unclustered field, given our observational setup.  The LBGs do not spatially cluster around either QSO (Fig.~\ref{j0338_pos}). In this field we also targeted a lower-priority object which was situated within a few arcsec of the QSO. Emission from the QSO increased the uncertainty on its photometry, contributing to it being formally assigned a lower priority. However, its  proximity to the QSO made it an intriguing candidate for spectroscopic follow up. Spectroscopy (Fig.~\ref{2a1spectra}) showed  a single, spatially-unresolved line at 7327 \AA\ coincident with the source position. The line is consistent with Ly$\alpha$ at $z=5.024$, within $\Delta z=0.003$ ($\Delta v$=150 kms$^{-1}$) of the QSO.  This emission line is highly unlikely to be due to extended Ly$\alpha$ emission from the QSO rather than Ly$\alpha$ emission from an independent source - firstly, because of the redshift offset between the two sources and secondly, because prior experience \citep{Bremer92,Heckman91} shows that extended QSO emission should be spatially resolved and fill at least part of the slit.

Within this field we find nine objects, including the two QSOs, that have redshifts separated by a few thousand km s$^{-1}$. Although the system is not likely to be viralised and the uncertainties are large given the small number of objects, the line-of-sight dispersion of the system can be estimated. We fit a Gaussian plus a constant model to the distribution of Ly$\alpha$ determined redshifts, in order to model the galaxies in the redshift spike and any foreground/background galaxies together. In this case the distribution is well fit by just a Gaussian, with almost no constant component to represent foreground/background galaxies, giving a line-of-sight dispersion of the galaxies in the redshift spike (the Gaussian component) of $900 ^{+1100}_{-400}$ kms$^{-1}$. The errors are the 1$\sigma$ errors of the fitting procedure, which dominate over the redshift errors.

\begin{table*}
    \caption{The spectroscopically confirmed z$\sim$5 LBGs and QSOs in the three fields. The error on the redshift is 0.001 for grade A LBGs and 0.003 for grade B LBGs. The redshifts of the J0338+0021, J1204-0021 and J2130+0026 QSOs from SDSS \citep{Schneider10}, which uses absorption lines to determined the redshifts, are 5.0319, 5.0319 and 4.951 respectively. In all cases we have determined the redshifts from the peak of the Ly$\alpha$ line and hence the slight discrepancy between the observed redshift and the published redshift for the two QSOs we targetted. *Spectroscopy of these objects was not obtained and the redshifts stated below are from the literature \citep{Djorgovski03,Schneider10}. $^{\dagger}$The close companion to QSO J0338 (see text). Its photometry may be contaminated by scattered light from the QSO. }
    \centering
    \begin{tabular}[t]{c c c c c c c c c}
    Slit Ref. & RA /deg & Dec /deg & Redshift & I /mag & R-I /mag & I-z /mag & Priority & Grade \\
    \hline
    J0338+0021 QSO  & 54.62211 & 0.36553 & 5.027 & 19.9$\pm$0.001 & 1.3$\pm$0.002 & 0.1$\pm$0.003 & 1 & A \\
    RD567 QSO* & 54.62513 & 0.31122 & 4.960 & 21.1$\pm$0.003 & 1.2$\pm$0.008 & 0.0$\pm$0.01 & na & na \\
    1a11 & 54.57256 & 0.40150 & 5.00  & 24.9$\pm$0.13 & 1.4$\pm$0.28 & 0.1$\pm$0.33 & 1 & B \\
    1a12 & 54.58068 & 0.40493 & 5.090 & 25.5$\pm$0.19 & 1.7$\pm$0.42 & 0.2$\pm$0.48 & 1 & A \\
    2a1$^{\dagger}$ & 54.62303 & 0.36554 & 5.024 & 25.7$\pm$0.46 & 1.2$\pm$0.14 & 0.1$\pm$0.30 & 2 & A \\
    2a6  & 54.65893 & 0.38288 & 5.027 & 24.8$\pm$0.13 & 2.0$\pm$0.39 & 0.0$\pm$0.33 & 1 & A \\
    2a11 & 54.64906 & 0.40444 & 4.987 & 24.7$\pm$0.13 & 2.1$\pm$0.39 & 0.4$\pm$0.24 & 2 & A \\  
    2b3  & 54.62834 & 0.32185 & 5.050 & 23.7$\pm$0.03 & 1.2$\pm$0.08 & 0.1$\pm$0.09 & 3 & B \\
    2b10 & 54.68122 & 0.35007 & 5.069 & 24.9$\pm$0.13 & 1.4$\pm$0.28 & 0.9$\pm$0.18 & 2 & A \\
    \hline 
    J1204-0021 QSO  & 181.17406 & -0.36364 & 5.086 & 19.2$\pm$0.0006 & 1.5$\pm$0.002 & -0.1$\pm$0.001 & 1 & A \\
    1b10 & 181.17325 & -0.38480 & 5.424 & 26.1$\pm$0.31 & $>$1.2$\pm$0.59 &  $<$0.6$\pm$0.55 & 1  & A \\
    1b14 & 181.18966 & -0.37056 & 4.977 & 24.8$\pm$0.13 & 1.4$\pm$0.20 & -0.1$\pm$0.24 & 1  & A \\
    2a1  & 181.18042 & -0.36293 & 5.101 & 25.5$\pm$0.20 & $>$1.8$\pm$0.54 &  0.0$\pm$0.50 & 1  & A \\
    2a3  & 181.18167 & -0.35564 & 5.050 & 24.9$\pm$0.13 & 1.5$\pm$0.26 &  0.7$\pm$0.18 & 2  & B \\
    \hline
    J2130+0026 QSO* & 322.53727 & 0.43612 & 4.951 & 20.5$\pm$0.004 &  0.9$\pm$0.005 & -0.3$\pm$0.009 & na & na \\ 
    1a5  & 322.50040 & 0.45221 & 4.777 & 25.2$\pm$0.15 &    1.0$\pm$0.22 & -0.4$\pm$0.45 & 2  & A  \\  
    1a8  & 322.50383 & 0.46411 & 5.133 & 25.4$\pm$0.23 & $>$1.8$\pm$0.44 &  0.0$\pm$0.44 & 1  & A  \\
    \end{tabular}
    \label{summary_0338}
   \end{table*}

\subsection{J1204-0021}
\begin{figure}
\includegraphics[width=0.95\columnwidth]{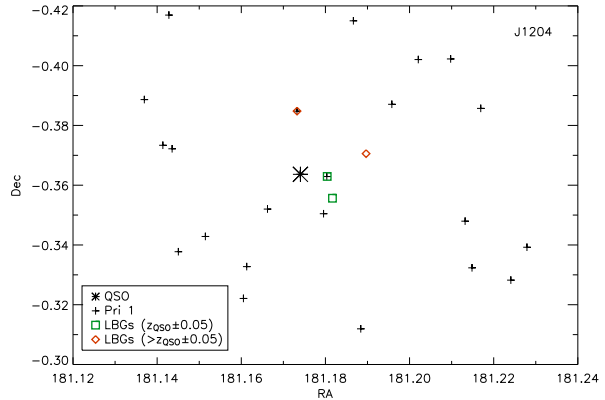}
\caption{The 2D spatial distribution of LBGs (diamonds) relative to the J1204-0021 QSO (star). The LBGs within $\Delta z$ = 0.05 of the QSO are shown with squares and those further away by diamonds. The priority 1 targets from photometry are shown in the background as crosses. }
\label{j1204_pos}
\end{figure}

Spectroscopy of the J1204-0021 field confirmed redshifts for four LBGs (three grade A, one grade B) from the high priority targets over the two masks. Of the four spectroscopically-confirmed LBGs, two fall within $\Delta z <0.04$ ($<2000$ kms$^{-1}$)  of the QSO redshift, a third  at $\Delta z =0.11$ and the fourth clearly  unconnected with the QSO at $\Delta z =0.34$.  The two LBGs with redshifts close to that of the QSO are  within 1 arcminute of it on the sky (Fig.~\ref{j1204_pos}). Again, from the ERGS survey,  we would expect zero or one object in a $\Delta z =0.1$ redshift bin for a field showing no sign of clustering. Thus while the significance of any clustering around the QSO in this field is obviously less than that for the J0338 field, confirming two objects close to the QSO (in combined spatial projection and redshift) implies some level of clustering around the QSO - a level that certainly could not be discerned from the numbers of LBG candidates selected from the photometry alone.

\subsection{J2130+0026}
\begin{figure}
\includegraphics[width=0.95\columnwidth]{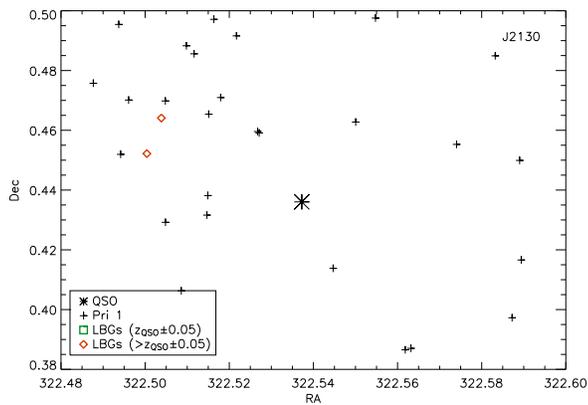}
\caption{The 2D spatial distribution of LBGs (diamonds) relative to the J2130+0026 QSO (star). The LBGs within $\Delta z$ = 0.05 of the QSO are shown with squares and those further away by diamonds. The priority 1 targets from photometry are shown in the background as crosses.}
\label{j2130_pos}
\end{figure}

Spectroscopy of the J2130+0026 field confirmed just two LBGs across the two masks. Neither were at redshifts close to that of the QSO (Fig.~\ref{j2130_pos}). The imaging of this field contained significant scattered light from nearby bright stars, the effects of which were only properly understood and dealt with after the spectroscopic masks had been designed. Once the effects of the scattered light were fully corrected for, the subsequent photometry indicated that up to eight extra high priority objects could have been placed on the mask. With a spectroscopic  confirmation rate of around 20\% for high redshift LBGs, this suggests up to 2 more LBGs may have been detected in this field. Therefore, it is possible that clustering similar to that seen in the J1204 field may be present here. Nevertheless, it is obvious that this field does not show the clear and strong clustering signature of the J0338 field.

\section {Discussion}

\subsection{Comparison to clustering in ERGS}

\begin{figure}
\centering
\includegraphics[width=0.95\columnwidth]{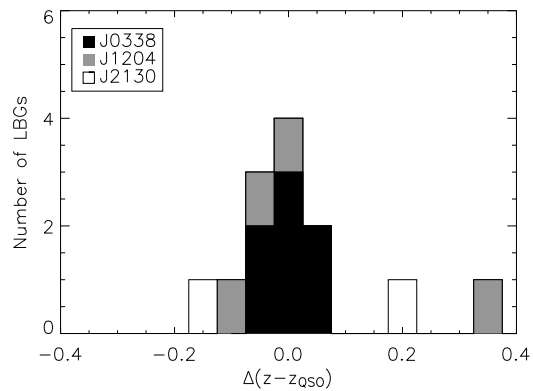}
\caption{The total number of LBGs found per 0.05 redshift bin away from the QSO. There is a clear spike at the 
redshift of the QSO but the majority of that signal comes from just one field, J0338. J2130 has no confirmed LBGs at 
the redshift of the QSO. }
\label{zdist_all}
\end{figure}

If we consider the spectroscopic results of the three fields together (see Fig.~\ref{zdist_all}), it is clear that LBGs do cluster around $z\sim5$ SDSS QSOs. However, the strength of the clustering signal varies from QSO to QSO.  Most of the signal in Fig.~\ref{zdist_all} comes from J0338 and none from J2130.  While this is consistent with the often contradictory results found in the literature for the properties of high redshift QSO environments \citep[e.g.][]{Stiavelli05,Zheng06,Kim09,Utsumi10}, the use of spectroscopy, along with the availability of an effective control sample from the ERGS survey, confirms that this variation is real rather than an effect of different observational strategies and QSO redshifts. The comparison we have made takes into account  both the completeness of the spectroscopic follow-up and the effectiveness of the spectroscopy, given that not every LBG can be spectroscopically confirmed with the quality of the data and the exposure times used.

In particular, we can compare the clustering of LBGs around QSOs to that found in the ERGS fields - which were not selected on the presence of a known distant object. Eight of the ten fields in ERGS show no sign of clustering in redshift and, at $z\sim 5$, show typically zero or one spectroscopically-confirmed source per $\Delta z=0.05$.  The same is true of the field targeted with the same observational setup and strategy in \citet{Lehnert03} and three other fields that surround it \citep{Davies11}. Additionally the three QSO fields in this work show no sign of any clustering at redshifts other than that of the QSOs. 

However, two fields in ERGS show significant spikes in their redshift distribution, one with seven objects between $5.11<z<5.21$ (J1040; see Fig.~\ref{cl1040}) and another with 17 objects between $4.95<z<5.15$ - most of these split into two narrow spikes in the redshift distribution (J1054-12; see Fig.~\ref{cl1054}). Plotting the spatial position of the objects in the two spikes in the J1054-12 field (Fig.~\ref{cl1054radec}) shows that they cluster spatially as well as in redshift. It appears that there are two separate structures in this field that may potentially merge to form a single larger one. If a representative $\sigma$ in km s$^{-1}$ is determined for each of the ERGS structures using the same method as for the spike in the J0338 field, similar values are obtained: $\sim 1300^{+2400}_{-590}$ km s$^{-1}$ for J1040 and $850^{+2000}_{-350}$ and $640^{+1070}_{-210}$ km s$^{-1}$ for the two peaks in the J1054-12 field. This indicates that the redshift clustering in these fields is comparable or stronger to that in the J0338 field - although we note that if the redshift spikes in J1054-12 were not able to be separated in redshift, then we would have a broader distribution like that in J0338. However, the richest of the ERGs fields, J1054-12, was followed up with five MOS masks allowing a highly complete survey of all high priority objects. As with our other QSO fields, the J0338 field was observed with just two masks, in this case allowing about half of the highest priority objects to be observed (see Table \ref{number}).  It is likely that if more masks were used the number of objects in the redshift spike would increase, potentially to the level of that seen in J1054-12.

\begin{figure}
\centering
\includegraphics[width=0.95\columnwidth]{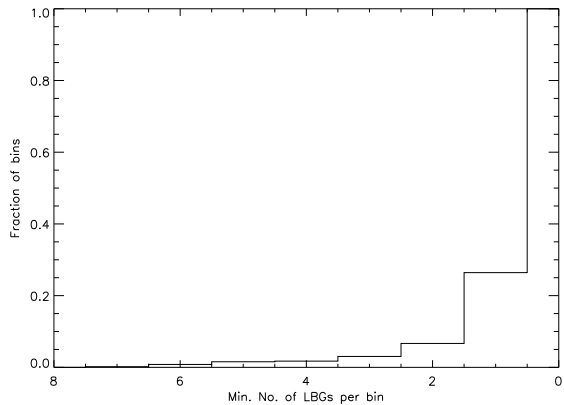}
\caption{The cumulative probability distribution of  finding $n$ or more (with $n$ running from zero to eight) LBGs in a $\Delta z=0.05$ redshift bin in the 14 FORS2 fields comprising the ERGS and BDF surveys. Most of the signal in the $n\geqslant 2$ bins arises from the two structures in Fig.~\ref{cl1040} and \ref{cl1054}.}
\label{frac_clustered}
\end{figure}

\begin{figure}
\centering
\includegraphics[width=0.95\columnwidth]{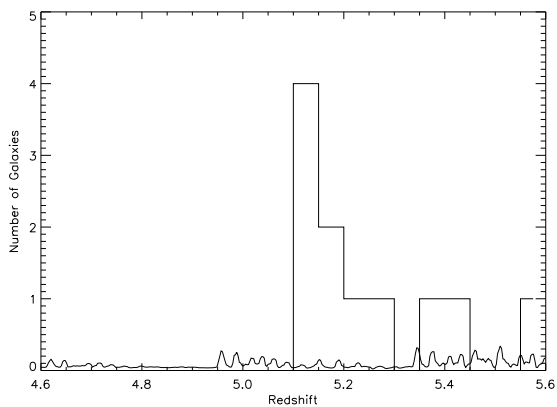}
\caption{A histogram of the distribution of spectroscopically confirmed LBGs found in the J1040 field of ERGS. Shown underneath is the spectrum of sky emission. }
\label{cl1040}
\end{figure}

\begin{figure}
\centering
\includegraphics[width=0.95\columnwidth]{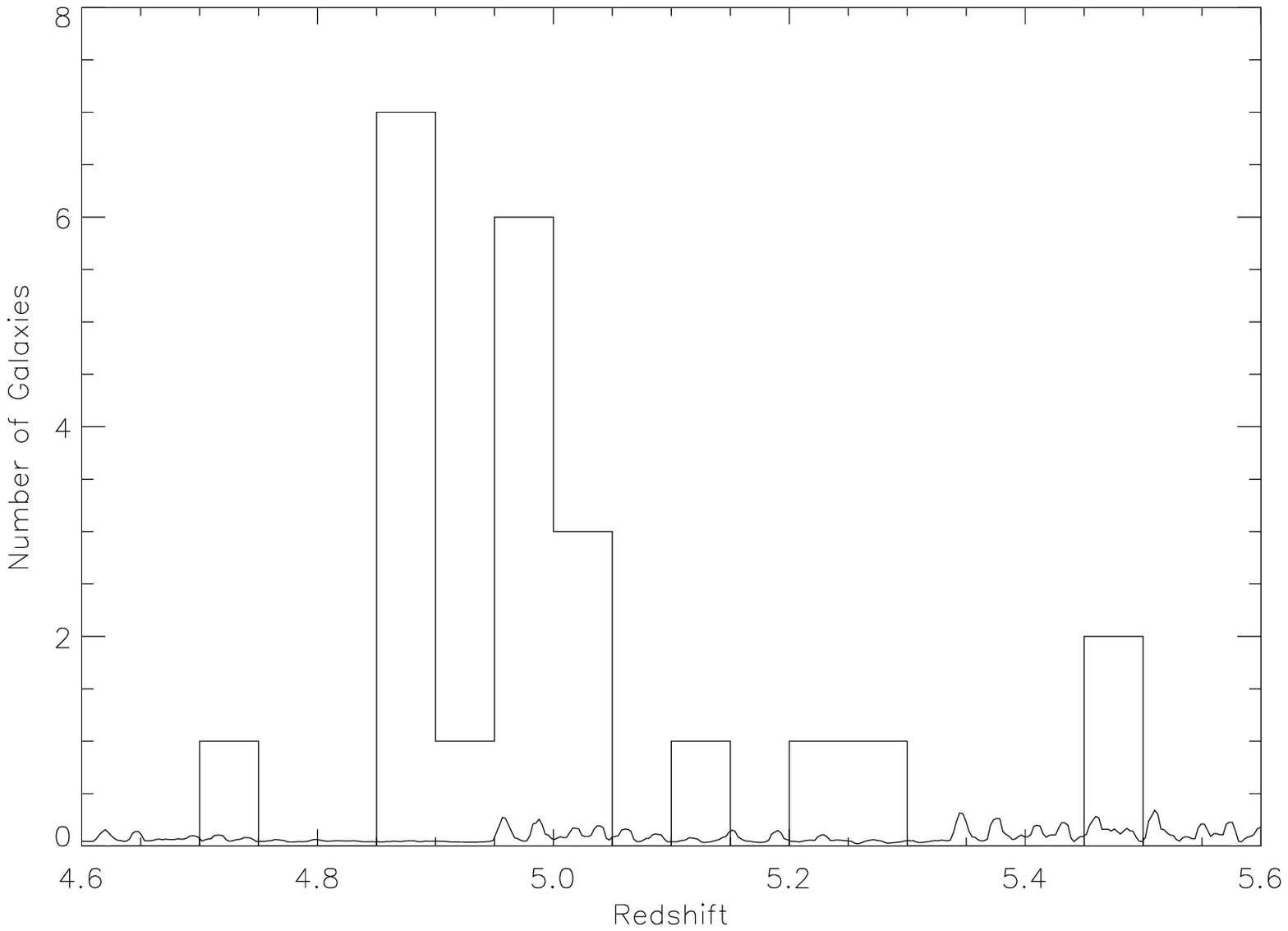}
\caption{A histogram of the distribution of spectroscopically confirmed LBGs found in the J1054-12 field of ERGS. Shown underneath is the spectrum of sky emission. }
\label{cl1054}
\end{figure}

\begin{figure}

\includegraphics[width=0.95\columnwidth]{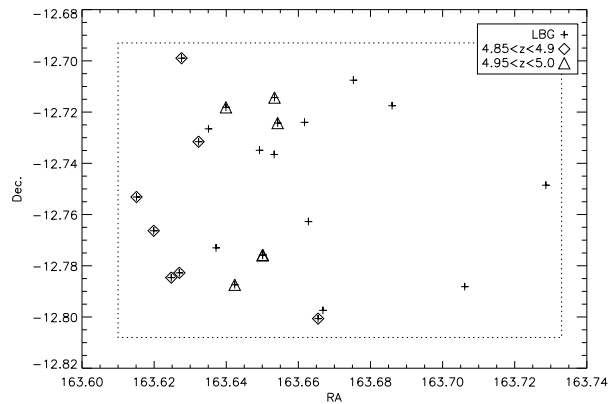}
\caption{The 2D spatial distribution of spectroscopically confirmed LBGs in the ERGS field J1054-12. Those in the two redshift spikes are highlighted as diamonds and triangles and the dotted lines shows the extent of the FORS2 field of view. }
\label{cl1054radec}
\end{figure}

Given these results, searching for high-redshift structures by targeting QSOs appears to be not much more efficient than observing blank fields. However, if we consider the statistics not in terms of fields, but in terms of independent redshift bins, the picture is subtly different. If we take all the `blank' sky fields, from the ERGS survey and the four fields in the area targeted by \citet{Lehnert03}, and randomly sample bins of width $\Delta$z = 0.05 between 4.8$<z<$5.3, we find only 7 per cent of such bins contain 2 or more LBGs. Similarly, 3 per cent of such bins contain 3 or more LBGs and 2 per cent contain 5 or more (see Fig.~\ref{frac_clustered}), with the majority of these bins arising from the two ERGS fields  showing spikes in their redshift distribution. Given the results presented here, where two of the three QSOs show evidence of objects at the same redshift, LBGs tend to cluster around $z\sim 5$ QSOs more often than they do around an arbitrary point at high redshift.

It appears that, as well as there being a clear variation in QSO to QSO environment marked by surrounding LBGs, even the strongest clustering around z $\sim$ 5 QSOs is no stronger than that seen in the field. Of the three QSO fields, one shows no evidence for clustering, another shows significant evidence for clustering of LBGs, whilst the third shows some evidence of clustering, but with only two LBGs at the QSO redshift.  Given that the probability of finding two LBGs (three if the QSO host is included) in a single random redshift bin not associated with a richer structure is so low in our previous work (Fig.~\ref{frac_clustered}), even two LBGs at the same redshift as a QSO indicates that the QSO is found in an over-dense structure.

\subsection{Clustering scales and markers}

Analysis of the Millennium-XXL simulation by \citet{Angulo12}  showed that the eventual fate of halos hosting high-redshift QSOs (specifically those at $z=6$) is most reliably determined from the richness of their environments on scales of 7-14 comoving Mpc. Those with the richest environments on these scales have the highest probability of ending up in massive clusters at the present epoch. Consequently, clustering on scales of a few Mpc may not be the strongest signature of a forming galaxy cluster. 

Here we note that the over-density of LBGs in the J0338 field is spread over (and probably extends beyond) the $\sim 7 \times 7$ arcmin$^2$ ($\sim 15 \times 15$ comoving Mpc$^{2}$) field, rather than clustering on smaller scales around just one of the QSOs. The over-densities in the ERGS fields are also spread over similarly large scales, as is the likely $z\sim 6$ protocluster identified by \citet{Toshikawa12}, rather than clustering on the 1-2 arcminute ($\sim$ 2 comoving Mpc) scales typical of more evolved clusters seen at $z\sim 1-2$ (3 Gyr later). Given the results of \citet{Angulo12}, it is likely that these over-densities mark out the first stages in the formation of massive clusters.

In our current analysis we have assumed that LBGs are reliable tracers of the underlying density distribution, and a lack of LBGs clustering at the redshift of the QSOs indicates that the QSO environment is not particularly over-dense. It is of course possible that we are being misled and that the number of detectable LBGs does not strongly correlate with the density field around a QSO. Given that galaxies selected through the Lyman break technique are simply those with significant unobscured rest-frame UV emission, arising from ongoing star formation (at rates of $\sim 10-30$ M$_\odot$yr$^{-1}$), it is unlikely that these are anything other than typical galaxies at their redshift. They should therefore trace the underlying density field as well as any other population which is undetected in the rest-frame UV to our flux limits.

It is possible that the presence of the QSOs may influence the colours and morphologies of galaxies in their immediate neighbourhoods due to an interaction between those galaxies and both the ionizing radiation field and matter outflows from the AGN. Such galaxies would  lie only a few arcseconds from the QSOs as we expect the AGN to be beamed in our direction, and for a galaxy to be caught within the beam it would need to be projected close to the QSO line of sight. Also it would need to be physically close to the QSO simply due to the drop-off with distance of the radiation field and outflow. We paid careful attention to all objects above our $I-$band magnitude cut in close proximity to the QSOs, both because of the possibility of interaction and because the proximity to the QSO could increase the uncertainty on the photometry. Where slit contention allowed we placed them on the masks e.g.~the close companion object to QSO J0338). Unless there is a population that is significantly reddened through interaction with a QSO (and therefore drops below our magnitude cut) it is unlikely that such objects were preferentially excluded from our spectroscopy.

Limited work has been carried out to explore the environments of high-redshift QSOs at other wavelengths, which might uncover redder or more obscured clustering populations.  \citet{Priddey08} found an excess of $S_{850\mu m}>4$ mJy sources within $\sim 1$  arcminute of three $z>5$ QSOs, indicating that dusty starbursts may cluster in their environments (two of the QSOs themselves were detected as hyperluminous dusty starbursts). One of the Priddey QSOs was previously studied by \citet{Stiavelli05}, who  identified tentative photometric evidence for an excess of distant LBGs in its immediate field. As yet none of the sources in the three fields studied by \cite{Priddey08} and \citet{Stiavelli05} have  published  spectroscopy  confirming the excess objects  to be at the redshift of their respective  QSO.

\subsection {Is J0338 typical of $z\sim 5$ SDSS QSOs?}

As most of the clustering signal seen in this work arises from a single QSO field, it is worth asking whether or not that QSO is typical of the $z\sim 5$ SDSS QSOs. Although not taken into account when this QSO field was selected, \citet{Djorgovski03} discovered a second fainter QSO (RD657) $\sim$ 3' away from SDSS J0338+0021 with a very similar redshift ($z=5.02$ for J0338, $z=4.96$ for RD657). They argued that this pair marked  a possible large-scale structure or protocluster. While there are no reported similar near-neighbours of the other two QSO fields probed in this work, there is no evidence in the literature that such neighbours have been searched for. However, \citet{Djorgovski03} note that of the 13 other $z>4.8$ QSOs studied by them, none appear to have a similar near-neighbour QSO. Therefore, it is reasonable to assume that the occurrence of QSO pairs on this scale is infrequent. Given the rarity of the QSO phenomenon, a rich environment clearly raises the probability of finding two QSOs in close proximity.

QSO J0338+0021 is also a strong  mm/submm source with continuum detections from 350$\mu$m to 1.2 mm, and detections of CO(5-4), [NII] and [CII] line emission \citep[summarised in][]{Wu09}, indicating  that the host galaxy is undergoing a strong ($\sim 2500$M$_\odot$yr$^{-1}$) starburst. Of the other two QSOs in this work, \citet{Carilli01} measured the flux of J1204-0021 to be $S=0.6\pm0.4$ mJy  at 250GHz (1.2 mm), six times fainter than J0338+0021, and no measurement of J2130+0026 at mm/submm wavelengths has yet been published.   

A subset of other high redshift QSOs have been observed at mm and submm wavelengths \citep[e.g.][]{Willott07, Priddey08}. Of those high redshift  QSOs observed at 1.2 mm \citep[e.g.][]{Carilli01,Bertoldi03,Petric03}, J0338+0021 appears to be one of the most luminous (within the top 15 per cent) indicating it is among a subset that are hyperluminous, with dust masses of $\sim 10^{8}$ M$_\odot$ or more and star formation rates of $>1000$ M$_\odot$ yr$^{-1}$. The presence of a starburst of this nature is a strong indication that the QSO host is particularly massive, and the more massive an early galaxy, the more likely it is to be found in a dense environment at high redshift \citep{Springel06}. Given the high SFR (and hence inferred mass) of this host galaxy,  the high sub-mm brightness of the source  may be a more important  influence over the strength of any surrounding over-density than the fact that it is a QSO. 

Little observational evidence for the nature of the environments of very distant hyperluminous sources currently exists. \citet{Capak11} identified a likely protocluster containing both a QSO and a strong sub-mm luminous starburst at $z\sim 5.3$, whilst observations of the two most over-dense ERGS fields at mm and sub-mm wavelengths resulted in no detections, although the full spatial extent of both fields was not probed \citep{Stanway08b,Stanway10,Davies10,Davies12}. However, our observations in the two ERGS fields are sensitive enough to detect any submm sources as bright as QSO J0338+0021 or indeed as bright as $\sim 40$ per cent of high redshift QSOs in \citet{Carilli01} etc., suggesting the presence of a sub-mm source is not essential for an over-density of this richness to form.

Inherent in the modelling of \citet{Angulo12} and other similar studies is an uncertainty in the fraction of massive high redshift structures that contain luminous QSOs. In the above we have assumed that the two structures identified in the ERGS fields differ from those around the QSOs in that they do not contain a powerful opticallly-luminous AGN. However, it is possible that they host QSOs that are beamed away from us, QSOs that are currently in their ``off" state or are otherwise obscured. If so, such sources may still be sub-mm bright objects even if the QSO emission is undetectable in the optical. Given our  existing mm and sub-mm data for these fields, we have no evidence that the two structures marked out by LBGs in the ERGS fields are centred on a misdirected or ``off" QSO.

 \section{Conclusions}
 
We have carried out an imaging and spectroscopic survey of the near ($\sim 6$ arcminute) environment of three $z\sim 5$ SDSS QSOs using a similar observational setup and selection to that used in ERGS. The similarity of approach confirmed that our strategy for finding LBGs is effective and provided examples of strong redshift clustering with which to compare the results in the QSO fields.
 
 We spectroscopically confirmed the presence of two or more LBGs at the same redshift as the QSOs in two of the three fields. Given that the probability of finding two or more LBGs in a narrow ($\Delta z \sim 0.05$) redshift range in 14 previously observed blank fields in ERGS and BDF is below 7 per cent, it is clear that QSOs do reside in over-dense regions relative to the field at $z\sim 5$. It is also evident that there is significant variation in  the strength of clustering of LBGs from QSO to QSO. Previous studies that relied on photometrically identifying over-densities of LBGs around QSOs  would have easily missed the more weakly clustered fields, where spectroscopy is needed to show that the QSO and LBGs are found in the same structure. Therefore, null results in these previous studies may have underestimated the richness of the QSO environments.
 
 Even the richest of the three QSO environments studied is no richer than the discrete structures identified in blank sky fields of ERGS. When all blank sky fields observed by us are taken into account, one such structure is detected per $\sim 7$ fields. If we split these  fields into redshift bins of width $\Delta z=0.05$, only $\sim 2$ per cent contain structures as rich as  that found in the richest QSO field. Hence, significant structure is more often found around QSOs than in `blank' sky regions. 
 
 We have been hesitant to call both the ERGS and the richest QSO-marked structures proto-clusters, as to do so requires evidence that the clustered regions will eventually form massive (M $>10^{14}$M$_\odot$) structures at lower redshift. However, by the standard of other claims for the discovery of such systems that exist in the current literature, these structures, with velocity dispersions of $<2000$ km s$^{-1}$, are very likely to be `proto-clusters'.

\section{Acknowledgments}
KH and LJMD acknowledge funding from STFC.  Based on observations made with ESO Telescopes at the La Silla and Paranal Observatory under programme IDs 082.A-0354, 083.A.-0618, 084.A.-0251, 085.A-0444 and 87.A-0589. The astronomical table manipulation and plotting software TOPCAT \citep{Taylor05} was used in the analysis of these data.

\bibliographystyle{mn2e}
\bibliography{bibliography}

\end{document}